\documentstyle[preprint,aps,prb]{revtex}
\tighten
\begin{document}
\draft
\title{A Cluster Method for the Ashkin--Teller Model}
\author{Shai Wiseman\cite{email1} and Eytan Domany\cite{email2}}
\address{Department of Electronics, Weizmann Institute of science, \\
	 Rehovot 76100 Israel }
\date{\today}
\newcommand{\av}[1]{\langle #1 \rangle}
\maketitle
\begin{abstract}

A cluster Monte Carlo algorithm for the Ashkin-Teller (AT) model
is constructed according to the guidelines of a general scheme for such
algorithms. Its dynamical behaviour is tested for the square lattice AT
model. We perform simulations on the line of critical points
along which the exponents vary continuously, and find that critical
slowing down is significantly reduced. We find continuous variation of
the dynamical exponent $z$ along the line, following the variation of the
ratio $\alpha/\nu$, in a manner which satisfies the Li-Sokal bound
$z_{cluster}\geq\alpha/\nu$, that was so far proved only for Potts models.

\end{abstract}

%\input epsf
% =-=-=-=-=-=-=-=-=-=-=-=-=-=-=-=-=-=-=-=-=-=-=-=-=-=-=-=-=-=-=-=-=-=-=-=-=
%\include{int}
\narrowtext
\section{Introduction}
The Ashkin-Teller (AT) model \cite{AT:1} has been studied extensively, ever
since its introduction, by a variety of methods. In two dimensions,
in particular, much is known about the phase diagram and critical
behavior of the model . Nevertheless there
are problems
that were not addressed extensively before; the critical behavior of random
AT models is one such issue about which not much is known, hope
for analytic treatment is slim, and therefore one expects numerical
simulations to be the main tool of investigation.
With this aim in mind, we set out to develop an efficient Monte Carlo
(MC) cluster algorithm for the AT model.

A convenient representation of the {\it generalized} AT model (gAT) is
in terms of two Ising spin variables, $\sigma_i$ and $\tau_i$, placed
on every site of a lattice. Denoting by $<ij>$ a pair of nearest
neighbor sites, the Hamiltonian is given by
$${\cal H}=-\sum_{<ij>} [
K_\sigma \sigma_i\sigma_j +
K_\tau \tau_i \tau_j +
L\sigma_i\sigma_j  \tau_i \tau_j]   \label{eq:AT1}$$
Here $K_\sigma$  ($K_\tau$) are the strengths of the interactions
between neighboring $\sigma$ ($\tau$) spins, and $L$ is a four-spin
coupling. The phase diagram of the ferromagnetic general AT model
is known in two dimensions
from duality transformations and renormalization group
studies \cite{Wu:Lin,Domany:Riedel}.  The three dimensional model has
been studied as well \cite{AT3d}.
In this paper we are concerned with the $Z(4)$ subspace of the
general model, in which $K_\sigma=K_\tau=K$. The phase diagram
in this subspace is reviewed in Sec. 2. The critical properties
of the model in this subspace are of special interest, since it
has a line of critical points, along which the exponents vary
continuously, and have been determined analytically \cite{Nienhuis},
interpolating between Ising and four-state Potts exponents.
For instance,
the value of the ratio $\alpha/\nu$
varies from 0 at the Ising ($L=0$) critical point to
$\alpha/\nu=1$ at the four-state Potts point $K=L$.
This exponent is of special interest to us since it has
been proved that for Potts models it serves as lower
bound to the dynamic exponent $z$ of cluster
algorithms \cite{Li:Sokal}.
Furthermore, the bound seems to be
reached in the case of the Ising and Potts models \cite{Li:Sokal}.

Cluster algorithms \cite{rev:clust} as introduced by Swendsen and Wang
\cite{SW:1}, and extended by Wolff \cite{Wol:1C}, are reviewed briefly
in Sec. 3. These algorithms give
rise to dynamics whose relaxation time, $\tau_{SW}$,
is significantly shorter than that of standard, single-spin flip
MC methods. This is most important
at a critical point, where
the relaxation time of a finite system grows with its linear
size L according to
$$\tau \sim {\rm L}^z \; .$$
Cluster algorithms have a significantly lower
dynamic exponent $z_{SW}$ than that of standard MC.
Hence if one is interested in performing extensive simulations
of a model such as AT, it is well worth to spend time on
developing an appropriate cluster algorithm.

Creating an efficient algorithm can be a challenge. Naive
application of the original SW scheme does not work (as
explained in Sec. 3 and demonstrated in Sec. 4). We set out to
construct an efficient cluster algorithm using the guidelines
and methods that were presented by Kandel and Domany \cite{DK:Gen} . This
general scheme is guaranteed to yield an algorithm that satisfies
detailed balance, and once the important excitations of the
model have been identified and incorporated, we are guaranteed to
get a good algorithm. This general formalism is briefly reviewed,
and the resulting algorithm is presented in Sec. 3.
Interestingly, the cluster algorithm found this way is
identical to what would have been obtained had we used
Wolff's Ising embedding method \cite{Wol:1C,embed:1}, as shown in the
Appendix. Numerical results
are given in Sec. 4; in particular, efficiency of our
algorithm is tested by
measuring the dynamic
exponent along the critical line, which
is indeed significantly lower than that of
standard MC.

An interesting question we set out to resolve concerns
comparison of $z_{SW}$ with $\alpha/\nu$ along the AT
critical line. We found that the Li-Sokal bound \cite{Li:Sokal}
$$z_{cluster} \geq {\alpha \over \nu}$$
is satisfied,
variation of the dynamic exponent follows that of $\alpha/\nu$,
and within our numerical accuracy and limitations due to finite
size effects, our results indicate that the two are equal.

\section{Phase diagram of the Ashkin Teller model}    \label{sec:AT}

 We now review the phase diagram of the square lattice AT model \cite{AT3d}
and some of its critical properties.
Figure \ref{fig:AT phase} gives the phase diagram of the model
 plotted as a function of the parameters $X$
and $Z$ where
\begin{equation}   \begin{array}{ll}
 Z=\exp^{-4K}\; ;   & X=\exp^{-2(K+L)} \;.
\end{array} \label{eq:def xz} \end{equation}
At times we use the notation $\vec{X}=(X,Z)$ to denote a point in phase
space. Full lines of phase transitions separate three phases:

(a) Paramagnetic, labeled as "P" . The couplings are sufficiently weak so
that the system is in a paramagnetic phase in which neither $\sigma$
nor $\tau$ (nor $\sigma\tau$) are ordered.

(b) Ferromagnetic phase, labeled as "F" . The couplings are sufficiently
strong so that $\sigma$ and $\tau$ independently order in a ferromagnetic
fashion so that $\langle\sigma\rangle=\pm\langle\tau\rangle$. In this phase
$\langle\sigma\tau\rangle$ is also different from zero and has the same sign
as $\langle\sigma\rangle\langle\tau\rangle$.

(c) A phase labeled $\langle\sigma\tau\rangle$, in which $\sigma\tau$ is
ordered
ferromagneticaly but $\langle\sigma\rangle=\langle\tau\rangle=0$. This phase
arises only for $L>K$.

On the dashed line $Z=X^{2}$ we have $L=0$ ; and thus it
 is obviously a subspace of two decoupled Ising models, having an Ising
transition at the point $X_{0}$.
 The dashed line $Z=X$ has $L=K$, in which case the AT model becomes
the 4-state Potts model. The point $X_{4}$ is a 4-state Potts multicritical
point.

A marginal operator
generates a continuous variation of critical exponents along the line
$X_{0}X_{4}$, isomorphic to
the known critical line of the eight-vertex model \cite{Baxter}.
Through an exact duality type transformation this line is mapped onto the
critical line of a staggered 8-vertex model and through a relation with the
Coulomb gas its critical exponents are known exactly \cite{Nienhuis}:
   \begin{equation}
y_{t}=2-(2/g_{R})
   \label{eq:Yt}  \end{equation}
where
   \begin{equation}
g_{R}=\frac{8}{\pi}\sin^{-1}[\frac{1}{2}\coth(2K)] \; ,
   \label{eq:gR}  \end{equation}
 and  $\frac{\gamma}{\nu}=\frac{7}{4}$  all along the line.
Lastly the lines $X_{4}B$ and $X_{4}C$ flow under renormalization to Ising
type fixed points.

% The transition line $X_{0}X_{4}$ is the self dual line $Z=1-2X$
%\cite{Domany:Riedel}. (cancel next subsection)
%-------------------------------------------------------------------------

The exact location of the transition line
$X_{0}X_{4}$ can be found through the duality transformation of the AT
model \cite{Domany:Riedel}. It is given by the self-dual line $Z=1-2X$.

%---------------------------------------------------------------------
\section{Cluster Method for the Ashkin Teller Model} \label{sec:AT clust}

\subsection{ SW cluster method}
 Cluster algorithms have proved to be a useful method of reducing critical
 slowing down in MC simulations. For completeness we review here the
pioneering
cluster algorithm  of Swendsen and Wang (SW) \cite{SW:1} for the Ising
model\cite{q} with
the Hamiltonian    \begin{equation}
  {\cal H}=-J\sum_{<k,j>}\sigma_{k}\sigma_{j}\;\;\;\;\;\;\;\;\sigma=\pm1\;.
   \label{eq:Is}  \end{equation}

The SW procedure stochastically identifies clusters of aligned spins,
and then flips whole clusters simultaneously.
 Starting from a given configuration $u$, SW go over all the bonds,
and either "freeze" or "delete" them.
A bond connecting two neighboring sites $k$ and $j$, is deleted with
probability $P_{d}$, and frozen with probability $P_{f}=1-P_{d}$, where:
   \begin{equation}
 P_{d}=e^{-\beta(J\sigma_{j}\sigma_{k}+J)} \; . \label{eq:swd1}
   \end{equation}
 Having gone over all the bonds, all spins which have a path of frozen bonds
connecting them, are identified as being in the same cluster. Now the
new configuration is generated by flipping every cluster with
probability 1/2 . Note that according to (\ref{eq:swd1}), only spins of the
same sign can be frozen in the same cluster.
 SW identify correctly the elementary large scale excitations as clusters of
 aligned spins. This correct identification is essential to the success of
a cluster method for any more general model.

In sec. \ref{sec:naive} we shall present a naive implementation of the
original SW scheme to the
AT model, and explain why it is not expected to work well. This necessitates
search for a different clustering rule; the one we found is based on a
general scheme which we now describe.

%-----------------------------------------------------------------------

\subsection{General scheme for cluster methods} \label{sec:general scheme}
A unifying view of all Cluster Algorithms has been given by Kandel and
Domany (KD)\cite{DK:Gen}, which we now review.
The general scheme consists of two steps: given a spin configuration $u$,
the first step consists of stochastically generating a new
Hamiltonian $\tilde{{\cal H}}$. The second step
consists of simulating the model with the new Hamiltonian, thereby bringing
it to a new configuration $u'$.
To carry out the first step, KD write the Hamiltonian as
    \begin{equation}
  {\cal H}=\sum_{l}V_{l}.
	 \end{equation}
 For example $V_{l}$ can be the energy of a single bond in the nearest
neighbor Ising Hamiltonian. Then to each $V_{l}$ they assign one of $N$
possible integers $i$, in a stochastic manner that
depends on the starting configuration $u$. That is, the probability to
assign $i$ to
$l$  is written as $ P_{i}^{l}=P_{i}^{l}(u)$. This probability is normalized,
i.e.
   \begin{equation}
  \sum_{i}P_{i}^{l}(u)=1
 \label{eq:norm}         \end{equation}
 for any term $l$ and configuration $u$. Then they construct a new
Hamiltonian:
   \begin{equation}
  \tilde{{\cal H}}_{\{i\}}=\sum_{l}\tilde{V}_{i(l)}^{l},
     \end{equation}
where (for any spin configuration $\tilde{u}$):
   \begin{equation}
\tilde{V}_{i}^{l}(\tilde{u})=V_{l}(\tilde{u})-\frac{1}{\beta}\ln[
     P_{i}^{l}(\tilde{u})]+C_{i}^{l}.    \label{eq:dbold}
     \end{equation}
The free parameters $C_{i}^{l}$ are configuration independent.

The second step consists of simulation of the model with any
procedure whose  transition probability $\tilde{T}_{\{i\}}(u\rightarrow
u')$ satisfies the detailed balance condition with respect to the new
Hamiltonian, i.e.
   \begin{equation}
  e^{-\beta\tilde{{\cal H}}_{\{i\}}(u)}\tilde{T}_{\{i\}}(u\rightarrow u')
= e^{-\beta\tilde{{\cal H}}_{\{i\}}(u')}\tilde{T}_{\{i\}}
(u'\rightarrow u). \label{eq:dbnew}
 \end{equation}
After completing the second step, a new configuration $u'$ is arrived at,
the original Hamiltonian is restored and the process is repeated.
Equations (~\ref{eq:dbold}) and (~\ref{eq:dbnew}) ensure that
the whole procedure satisfies the detailed balance condition, with respect
to the original Hamiltonian (for
the proof see \cite{DK:Gen} ), but ergodicity needs to be proved for each
application separately.

We give now two types of modifications to the Hamiltonian.
Consider a term $V_{l}(u)$ that can take $M$
distinct energy values $E_{i}$, $i=1$,\ldots,$M$.
The first is the deletion operation, used by SW, which eliminates the
interaction $V_{l}(\tilde{u})$ that gets replaced by
   \begin{equation}
\tilde{V}_{d}^{l}(\tilde{u})=0, \label{eq:del int}
     \end{equation}
for any configuration $\tilde{u}$. To get this, we must have ( see eq.
(~\ref{eq:dbold})):    \begin{equation}
P_{d}^{l}(u)=e^{\beta[V_{l}(u)+C_{d}^{l}]}.  \label{eq:del prob}
     \end{equation}
The constant $C_{d}^{l}$ must be chosen so that $P_{d}^{l}(u)\leq 1 $ for
any $u$.

The second modification is a {\em `generalized' freezing} operation which
we
will later use in our scheme for the Ashkin-Teller model. Its probability is
   \begin{equation}
P_{g}^{l}(u)=\left\{ \begin{array}{ll}
 e^{\beta[V_{l}(u)+C_{g}^{l}]}   & \mbox{if } V_{l}(u)=E_{m}\mbox{ and }
  1\leq m \leq \mu \\
0 & \mbox{if } V_{l}(u)=E_{m}\mbox{ and } \mu< m \leq M
   \end{array} \right. ,\label{eq:gf prob}\end{equation}
and the modified interaction according to (\ref{eq:dbold}) is
   \begin{equation}
\tilde{V}_{g}^{l}(\tilde{u})=\left\{ \begin{array}{ll}
   0 & \mbox{if } V_{l}(\tilde{u})=E_{m} \mbox{ and }1\leq m \leq \mu \\
   \infty & \mbox{if } V_{l}(\tilde{u})=E_{m}\mbox{ and } \mu< m \leq M
   \end{array} \right. .\label{eq:gf V}\end{equation}
Again the constant $C_{g}^{l}$ must be chosen so that $P_{g}^{l}(u)\leq 1 $
for any $u$. This type of modification  allows free movements between some
of the states $u$ of $V_{l}(u)$ which didn't have the same energy in the
original Hamiltonian, and we will use it for the Ashkin Teller model in
section \ref{sec:ATC}. Perhaps it can be useful in general, in cases where
the $V_{l}$s can have more than two possible energies. This type of
operation has been used for example in \cite{Lana}.

A particular case of this operation is obtained if $\mu=1$, and we set
   \begin{equation}
 C_{1}^{l}=\frac{1}{\beta}\ln(p_{1})-E_{1}.
   \end{equation}
 Substitution into (\ref{eq:gf prob}) yields
   \begin{equation}
P_{1}^{l}(u)=\left\{ \begin{array}{ll}
   p_{1} & \mbox{if } V_{l}(u)=E_{1} \\
   0 & \mbox{otherwise} \end{array} \right. ,
  \label{eq:freezing int.} \end{equation}
and from (\ref{eq:gf V}),
the modified interaction takes the form :
   \begin{equation}
\tilde{V}_{1}^{l}(\tilde{u})=\left\{ \begin{array}{ll}
   0 & \mbox{if } V_{l}(\tilde{u})=E_{1} \\
   \infty & \mbox{otherwise} \end{array} \right. .
  \label{eq:freezing prob} \end{equation}
This operation assigns infinite energy to any configuration $\tilde{u}$ for
which $V_{l}(\tilde{u})\neq E_{1}$. That is, in the ensuing simulation the
interaction $V_{l}$ is frozen at energy $E_{1}$. Freezing of $V_{l}$ is
assigned with probability $p_{1}$, and only when $V_{l}(u)=E_{1}$. It is
easy to see that the SW freezing is precisely of this form.

%----------------------------------------------------------------------

\subsection{Cluster Method for the Ashkin Teller model}\label{sec:ATC}
We describe now the cluster algorithm we devised for the
gAT model and
the considerations that lead us to it. It consists of a
freeze-delete scheme which generates non-interacting clusters of aligned
$\sigma$ spins and of aligned $\tau$ spins. We will phrase it in terms
of the general scheme described in sec. \ref{sec:general scheme}.

The first decision one needs to make when coming to design a freeze-delete
scheme is the choice of the basic interaction term $V_{l}$. Our choice
 is to associate all interactions that reside on an edge $<jk>$ of the
lattice to one $V_{l}$ :
\begin{equation}
V_{l}=-[K_{\sigma}\sigma_{k}\sigma_{j}
+K_{\tau}\tau_{k}\tau_{j}+ L\sigma_{k}\tau_{k}\sigma_{j}\tau_{j}] .
  \label{eq:Vl}  \end{equation}
 Since the model is invariant under any permutation of $K_{\sigma},K_{\tau}$
 and $L$ ( to make this symmetry explicit, define a new Ising spin
$s_{j}=\sigma_{j}\tau_{j}$ along with the
constraint $s_{j}\sigma_{j}\tau_{j}=1$ ), it is possible to choose
$L\leq K_{\sigma},K_{\tau}$ (the reason for this choice will become clear
later).
The interaction $V_{l}$ depends on
four independent Ising spins that can have
 16 states. Every $V_{l}$ can take one of four possible energy values:
 \begin{equation}  \begin{array}{l}
 E_{1}=-K_{\sigma}-K_{\tau}-L   \\
 E_{2}=K_{\sigma}-K_{\tau}+L   \\
 E_{3}=-K_{\sigma}+K_{\tau}+L   \\
 E_{4}=K_{\sigma}+K_{\tau}-L   \\
 \end{array}.\end{equation}
Every energy is four-fold degenerate; we denote by $u_{i}$ all (four) states
for which $V_{l}=E_{i}$. Let $u_{i}$ represent the state of the spins
$\sigma_{j},\;\sigma_{k},\;\tau_{j},\;\tau_{k}$  of a particular  pair of
n.n. sites $<j,k>$.
 Four representative possible states are depicted in
figure \ref{fig:states}.
$u_{1}$ is a ground state in which both the $\sigma$ and
the $\tau$ bonds are satisfied. $u_{2}$ and $u_{3}$ are excited states in
which
one Ising bond is broken  and the other is satisfied.
 Our choice $L\leq K_{\sigma},K_{\tau}$ makes $u_{4}$ the highest
energy state\cite{L<K}, in which both Ising bonds are broken.

The general philosophy of our freeze-delete scheme is as follows. We wish to
build clusters of $\sigma$ spins and of $\tau$ spins since we know that
these clusters are the basic excitations of the model. Clusters of
$s=\sigma\tau$
spins are not important because $L$, the bond between them, is the
weakest.
In order to build
clusters,  one needs to freeze parallel $\sigma$ spins to each other and
parallel $\tau$ spins to each other, and delete the bonds between
antiparallel spins. For example in the state $u_{2}$ we wish to freeze the
bond between the $\tau$ spins and delete the bond between the $\sigma$
spins. This consideration leads us to include in our scheme
two operations. The operation which we will identify as i=2 freezes the
bond between $\tau_{j}$ and $\tau_{k}$ and deletes the bond between
$\sigma_{j}$ and $\sigma_{k}$ (see fig. \ref{fig:scheme} ). According to the
discussion
above, we want to perform this operation with some probability $p_{2}\not=0$
 for the state $u_{2}$, and with probability 0 for the states $u_{3}$ and
$u_{4}$. This operation allows one to move from $u_{2}$ to $u_{1}$, so in
order to maintain detailed balance, we must perform it with some
non vanishing probability
$q_{2}\not=0$ on $u_{1}$ too. All this is achieved by assigning, in the
modified
interaction term $\tilde{V}_{2}^{l}$, infinite energy to $u_{3},u_{4}$ and
zero energy to $u_{1},u_{2}$. This modification is precisely of the form of
eq. (\ref{eq:gf V}) so it is of the generalized freezing type described
at the end of section \ref{sec:general scheme} and
hence the probabilities must be (see eq. (\ref{eq:gf prob}))\cite{coup}:
 \begin{equation} P_{2}^{l}(u)=\left\{ \begin{array}{ll}
 e^{V_{l}(u)+C_{2}}   & \mbox{if } V_{l}(u)=E_{2}\mbox{
or  } V_{l}(u)=E_{1}   \\
0 & \mbox{if } V_{l}(u)=E_{3}\mbox{  or  } V_{l}(u)=E_{4}
   \end{array} \right. .\label{eq:i=2 prob}\end{equation}
 The choice of $C_{2}$ will set the values of $p_{2}$ and $q_{2}$ :
   \begin{equation} \begin{array}{ll}
 p_{2}=e^{V_{l}(u_{2})+C_{2}} & q_{2}=e^{V_{l}(u_{1})+C_{2}} \;.
   \end{array}  \label{eq:p2q2}\end{equation}

The operation which we identify as i=3 freezes the bond between $\sigma_{j}$
 and $\sigma_{k}$ and deletes the bond between $\tau_{j}$ and $\tau_{k}$. It
follows the same logic as the operation i=2 and its probability is:
\begin{equation}
P_{3}^{l}(u)=\left\{ \begin{array}{ll}
 e^{V_{l}(u)+C_{3}}   & \mbox{if } V_{l}(u)=E_{3}\mbox{
or  } V_{l}(u)=E_{1}   \\
0 & \mbox{if } V_{l}(u)=E_{2}\mbox{  or  } V_{l}(u)=E_{4}
   \end{array} \right. \label{eq:i=3 prob}\end{equation}
  Again we define the probabilities:
   \begin{equation} \begin{array}{ll}
 p_{3}=e^{V_{l}(u_{3})+C_{3}} & q_{3}=e^{V_{l}(u_{1})+C_{3}}  \;.
   \end{array}  \label{eq:p3q3}\end{equation}

 The third operation we wish to perform is a freezing operation of the form
defined in section \ref{sec:general scheme} equations (\ref{eq:freezing
int.}-\ref{eq:freezing prob}). In our case it freezes
 the bond between $\sigma_{j}$
 and $\sigma_{k}$ {\em and} the bond between $\tau_{j}$ and $\tau_{k}$,
so it
also builds the $\tau$ and $\sigma$ clusters. We'll denote it by i=1 and its
probability is:
  \begin{equation}
P_{1}^{l}(u)=\left\{ \begin{array}{ll}
   p_{1} & \mbox{if } V_{l}(u)=E_{1} \\
   0 & \mbox{otherwise} \end{array} \right. .
  \label{eq:freezing AT} \end{equation}

 The fourth and last operation is the deletion operation defined in section
 \ref{sec:general scheme} equations (\ref{eq:del int},\ref{eq:del prob})
  with the probability:
   \begin{equation}
P_{d}^{l}(u)=e^{V_{l}(u)+C_{d}}.  \label{eq:del AT}
     \end{equation}

These four operations define completely our freeze-delete scheme, summarized
by fig. \ref{fig:scheme}. In the first column, in each row, one or two
representative configurations of $u_{i}$ appear,depicting the state of
$\sigma_{j},\tau_{j},\sigma_{k}$ and $\tau_{k}$.
 In the first row the four modified interactions $\tilde{V}_{i}^{l}$
 are depicted. An upper double line for example, denotes a
frozen bond between $\sigma_{j}$ and $\sigma_{k}$, a blank denotes a deleted
bond.
We have yet to determine the constants $C_{2},C_{3},
 p_{1},C_{d}$. In the case of $u_{4}$, we want to delete the
bond between the $\tau$ spins and the bond between the $\sigma$
spins with probability 1, as we never want an unsatisfied bond to be
frozen, so we choose $C_{d}=-E_{4}$, to get $P_{d}(u_{4})=1$.
Having chosen $C_{d}$, all deletion probabilities are set.
The rest of the constants are determined by the normalization condition
(\ref{eq:norm}), i.e.
 \begin{equation}     \begin{array}{l}
 p_{2}=1-P_{d}(u_{2})  \\
 p_{3}=1-P_{d}(u_{3})   \\
 \end{array}. \label{eq:norm1 AT} \end{equation}
Consequently, equations (\ref{eq:p2q2}, \ref{eq:p3q3}) determine the
constants $C_{2},C_{3}$ and the probabilities $q_{2},q_{3}$. Finally,
$p_{1}$ is determined again by the normalization condition
 \begin{equation}
 p_{1}=1-P_{d}(u_{1})-q_{2}-q_{3} .
 \label{eq:norm2 AT} \end{equation}
For completeness we'll  list all the probabilities of our scheme which
follow from equations (\ref{eq:i=2 prob}--\ref{eq:norm2 AT}):
 \begin{equation}  \begin{array}{ll}
 P_{d}(u_{i})=e^{E_{i}-E_{4}} &\;\;\;\;\;\forall\mbox{i}  \\
 p_{i}=1-e^{E_{i}-E_{4}} &\;\;\;\;\;\mbox{i}=2,3   \\
 q_{i}=p_{i}e^{E_{1}-E_{i}} &\;\;\;\;\;\mbox{i}=2,3   \\
 p_{1}=1-P_{d}(u_{1})-q_{2}-q_{3}&  \\
 \end{array}.\label{eq:prob AT} \end{equation}
 Checking that all the probabilities of the scheme fulfill the condition
$0<P_{i}(u)<1$ for all $i$ and $u$ is trivial for all probabilities except
for $p_{1}$ which is a bit more laborious, but still straightforward.

We've described how we generate a new Hamiltonian of non-interacting
clusters of
$\sigma$ spins and of $\tau$ spins. Now we have to choose some legitimate
MC procedure to simulate this Hamiltonian. We chose to do it in a similar
fashion to the single cluster algorithm of Wolff \cite{Wol:1C}, but the SW
version
is equally applicable. To be more explicit, we choose a site $j$ of the
lattice at random and a random spin, either $\sigma_{j}$ or $\tau_{j}$ (
The choice of $\sigma$ or $\tau$ was done with probability 1/2 which is
sensible in the case $K_{\sigma}=K_{\tau}$, but in general any probability is
acceptable, and an optimal choice can be made to minimize the
autocorrelation time $\tau$). This
spin will belong to the cluster we will flip, and the cluster will contain
only $\tau$ ($\sigma$) spins if we initially chose $\tau_{j}$ ($\sigma_{j}$).
We perform the freeze-delete
operations only on bonds belonging to the surface of the cluster. For
example suppose at some stage $\tau_{j}$ was joined to the cluster, and
suppose that the term $V_{j,k}$ was modified according to i=3, then
$\tau_{k}$ will not be joined to the cluster and we need not perform
freeze-delete operations on the other three bonds connecting the site $k$.
If it is modified according to i=1 or i=2 then $\tau_{k}$ is joined to the
cluster. Since our cluster algorithm fits into the general scheme, we do not
need to prove detailed balance, while ergodicity is ensured by the deletion
operation.

  We've explained the reasoning behind our algorithm. We believe that
the clusters we build of parallel $\sigma$ spins and of parallel $\tau$ spins
are the basic excitations of the model. We therefore believe that it will
be efficient. It is encouraging to notice that in the decoupled Ising
subspace and the 4-state Potts subspace our freeze/delete scheme is
identical with SW's freeze/delete operations for these models. In the
appendix we show that our algorithm is equivalent to the idea of embedding
 into the AT model an Ising model and simulating it by Wolff's single
cluster procedure.
 In sec. \ref{sec:res} we present numerical evidence for our
algorithm's efficiency.
%---------------------------------------------------------------------------

\subsection{`Naive' SW option}  \label{sec:naive}
 We find it illuminating to compare our algorithm to a cluster algorithm
which one could regard as the naive generalization of the SW method to the
AT model. Such an algorithm would define $V_{l}$ in the same manner as our
scheme does ( see eq. \ref{eq:Vl}).
For each $u_{i}$, the bonds between the two neighboring sites get either
deleted ( with our $P_{d}(u_{i})$ ) or frozen with $P_{f}(u_{i})=1-
P_{d}(u_{i})$. Fig. \ref{fig:SW AT} can clarify how this scheme fits into
the general one of sec. \ref{sec:general scheme} and how it compares
with our scheme. Since it fits into the general scheme we do not need to
prove detailed balance, while the deletion operation ensures its ergodicity
except for $T=0$.

 This scheme also generates clusters of $\sigma$ spins
and clusters of $\tau$ spins, but
with the `naive' scheme, antiparallel Ising spins can be found in the same
cluster, so this scheme does
not identify the elementary excitations of the model. Moreover the $\sigma$
clusters' structure is forced to match the $\tau$
clusters ( and vice versa). This is totally unphysical since in practice
it is energetically favorable for the two cluster structures
\underline{not} to match ( for $L<K_{\sigma},K_{\tau}$ ).
 At $T=0$ this scheme could freeze the whole lattice into a single
cluster even when it is not in the ground state. This is in contrast to the
ability of our scheme to relax excitations on any scale even at $T=0$ ( in
other words, At $T=0$ our algorithm will bring any
initial configuration to the ground-state in a short time).
At a finite temperature the `naive' SW scheme will produce clusters that are
too large.
In sec. \ref{sec:MC res} we list results of simulations using the
single cluster (1C) version of this `naive' SW scheme, which show
that it is indeed much less efficient than our algorithm.
%---------------------------------------------------------------------------

%\include{resb}

\section{Simulations of the AT critical line}   \label{sec:res}
In addition to checking the efficiency of our method at the AT critical
line,
we also wanted to check a prediction made by Li and Sokal \cite{Li:Sokal},
 who have proved a rigorous lower bound
\begin{equation}
z_{SW}\geq \alpha/\nu
 \label{eq:Li}   \end{equation}
 for the dynamical critical exponent of the SW
algorithm for the ferromagnetic q-state Potts model.
 More precisely, the Li-Sokal bound is $\tau_{exp/int}\geq{\mbox const}
\times C_{h}$, from which eq. (\ref{eq:Li}) follows
(for definitions of
$\tau_{exp/int}$ and $z_{exp/int}$ see \cite{Sokal:Lat}).
  This bound relates
dynamics to the static properties of a model and is thus of great
importance. Their proof is for
both $z_{int}$ and $z_{exp}$. We wanted to check whether this bound holds all
along the AT critical line, which connects the decoupled Ising critical point
with $\frac{\alpha}{\nu}=0$ at one end to the 4-state Potts critical
 point with $\frac{\alpha}{\nu}=1$ at the other end. At both of
these points our algorithm is identical to Wolff's 1C version of the SW
method  for Potts models. The Li-Sokal bound was proved for the SW
dynamics, but perhaps it is valid for Wolff's 1C dynamics as well (at
least for $d=2$). We
are unaware of any rigorous proof for that, but our results for $z$ for
$q=2$ and 4
seem to indicate $z_{1C}=z_{SW}$ and are in accordance with previous results
for the Ising critical point (see \cite{SW:1} and \cite{Wol:comp}).
 Besides estimating the dynamical exponents, we've estimated, using finite
size scaling, the critical exponents $\frac{\alpha}{\nu}$ and $\frac{\gamma}
{\nu}$ along the AT critical line.
%The AT critical line can be found by duality as we described in sec.
%\ref{sec:duality} and is given by
As stated in sec. \ref{sec:AT} the AT critical line is given by
\begin{equation}
 Z=1-2X.    \label{eq:AT fixed}
\end{equation}
Our measurements were done at the decoupled Ising critical point
$\vec{X}_{0}$ and at the 4-state Potts critical point $\vec{X}_{4}$ (see
 figure \ref{fig:AT phase} and exact definition in sec. \ref{sec:AT}.
Three additional measurements were
carried out at three intermediate equidistant points in the $X-Z$ plane, on
the AT critical line.  So a total of 5 measurements were done at the points
\begin{equation}   \begin{array}{ll}
 \vec{X}_{i}=\vec{X}_{0}+\frac{i}{4}(\vec{X}_{4}-\vec{X}_{0})&
i=0\ldots4 . \end{array} \label{eq:Xplace} \end{equation}
 All the points $\vec{X}_{i}$ are marked in fig. \ref{fig:AT phase} . We
simulated
lattices with periodic boundary conditions of up to size $128\times128$,
and up to $5\times10^{5}$ clusters were flipped for each lattice size.

We calculated the energy\cite{energy} per site:
 \begin{equation} E=\langle E\rangle=-\frac{1}{{\rm L}^{2}}\langle
\sum_{<k,j>}[K\sigma_{k}\sigma_{j}+K\tau_{k}\tau_{j}+
L\sigma_{k}\tau_{k}\sigma_{j}\tau_{j}] \rangle
    \label{eq:E}\end{equation}
where ${\rm L}$ is the linear lattice size,
and the angular brackets denote the usual thermal MC average.
 The specific heat per site follows from the energy fluctuations
\begin{equation}
C=\frac{C_{h}}{k_{B}}={\rm L}^{2}(\langle E^{2}\rangle-\langle E\rangle^{2})
.     \label{eq:Cv}\end{equation}
We calculated the magnetic susceptibility of the $\sigma$ spins defined
as\cite{energy}
  \begin{equation}
\chi={\rm L}^{2}\langle M^{2}\rangle\;\;\;\;\;\;\;\;M=\frac{1}{{\rm L}^{2}}
\sum_{k}\sigma_{k}  \label{eq:Xi}  \end{equation}
and also measured $c$, the size of the cluster flipped at each step, and
calculated $\langle c \rangle$. There is a connection between the size
of the clusters and the susceptibility \cite{Wolff:estimator} which
is common for algorithms which build non-interacting clusters of spins
(if all spins in a cluster have the same value). For
1C algorithms it has the simple form \cite{Wol:1C}
\begin{equation}
\chi=\langle c\rangle   .
 \label{eq:improved estimator}  \end{equation}
To get the dynamic properties we calculated the time dependent
autocorrelation functions $\phi_{E}(t)$ and $\phi_{\chi} (t)$  defined as
\begin{equation}
 \phi_{A}(t)=\frac{\langle A(0)A(t) \rangle-\langle A \rangle^{2}}
{\langle A^{2} \rangle-\langle A \rangle^{2} }\;,   \label{eq:fi(t)}
   \end{equation}
   where $A$ stands for $E$ or $\chi$.
A typical plot of $\phi_{\chi}(t)$, measured from the $X_{2}$ model at ${\rm
L}=128$ is presented in fig. \ref{fig:fi(t)}.

%---------------------------------------------------------------------------

\subsection{MC results and discussion} \label{sec:MC res}
\subsubsection{Susceptibility and Specific
heat} According to finite-size scaling theory \cite{Barber} one expects:
\begin{equation} \begin{array}{ll}
\chi\sim {\rm L}^{ \frac{\gamma}{\nu} } &\;\;\;\; C\sim {\rm L}^{ \frac{
\alpha}{\nu} }    \end{array} \label{eq:scale} \end{equation}
 for large enough L.
 Fitting our measurements of $\chi$ to eq. (\ref{eq:scale}) for
lattice sizes ${\rm L}\geq16$
fits the exact universal value $\frac{\gamma}{\nu}=\frac{7}{4}$ within
errors (see table \ref{tab-exponents}) as expected.
 Our measurements confirm the equality (\ref{eq:improved
estimator}), where $\langle\chi\rangle$ and $\langle c\rangle$ have an error
of the same magnitude.
 Plots of $\log C$ vs. $\log {\rm L}$ for the five models can be seen in
fig. \ref{fig:Cv graph}. From linear fits to the $\log-\log$ plots,
estimates for  $\frac{\alpha}{\nu}$ are obtained which do not agree with the
 exact known values. For comparison with the exact values see table
\ref{tab-exponents} and fig. \ref{fig:compZXiCv}. The differences may be due
to finite size effects and corrections to scaling . An exception to this
mismatch is the decoupled Ising
point for which the value of $\frac{\alpha}{\nu}=0$ fits nicely according to
the semi-log plot in fig. \ref{fig:TX0s} (for completeness we also quote in
table \ref{tab-exponents} an estimate for $\frac{\alpha}{\nu}$ at $X_{0}$ from
the
$\log-\log$ fit of fig. \ref{fig:Cv graph}). The slope of the $C$ curves at
$X_{1}$ and $X_{2}$  does tend to a lower value with increasing lattice size.
 The 4-state Potts model is known to have
 a logarithmic correction to scaling: $C\sim {\rm L}/\log^{3/2}{\rm L}$
\cite{Li:Sokal};
 in any case, our result $\frac{\alpha}{\nu}=.747(3)$ agrees with
previous MC results obtained from lattices of sizes up to ${\rm L}=256$
\cite{Li:Sokal}.

%---------------------------------------------------------------------------

\subsubsection{$\tau_{exp}$ and $\tau_{int}$}
In order to check the efficiency of our algorithm and to see whether the
Li-Sokal bound holds, we measured $\tau_{exp}$ and $\tau_{int}$ for both the
energy $E$ and the susceptibility $\chi$. This was done by the
following procedure. $\phi_{E/\chi}(t)$ was plotted on a semi-log plot, the
unit of time being a single cluster flip. $\tau_{exp}$ was then
consistently estimated by the slope of $\ln \phi(t)$ in a window from
$\tau_{exp}$ to $3\tau_{exp}$. A typical example for $ \phi(t)$ and the
linear fit to extract $\tau_{exp}$ can be seen in figure \ref{fig:fi(t)}.
To calculate $\tau_{int}$, we integrated numerically
$\phi(t)$ in the interval from $t=0$ up to $t\approx 1.5 \tau_{exp}$
 and
estimated the tail ($t>1.5 \tau_{exp}$ ) of $\phi(t)$ by the exponential
fit. The error of both
$\tau$'s was estimated from repeated experiments when the statistics were
large enough and otherwise from subjective estimates from the fluctuations
in $\tau_{exp}$ in the window $\tau_{exp}-3\tau_{exp}$.
Each MC step, or cluster flip, involves on the average the flipping of
$\langle c\rangle$ spins.
Since the natural unit of time is a sweep over all spins of the lattice, we
 multiplied $\tau$ by $\langle c\rangle{\rm L}^{-2}$.

 In figures \ref{fig:TX0l} through \ref{fig:ZcompXi} we plot measured
values
 of $\tau_{exp}$ and $\tau_{int}$ for  $E$ and $\chi$ as a function of the
 linear lattice size ${\rm L}$. In figures \ref{fig:TX0l}--\ref{fig:TX3}
we display the measurements of $\tau_{\chi}$ at different models
$X_{i}$ separately, while in figures \ref{fig:TexpE} and \ref{fig:ZcompXi}
we display $\tau_{exp,E}$ and $\tau_{int,E}$ for the five models jointly.
 Fitting our results to the forms
   \begin{equation} \begin{array}{ll}
   \tau_{exp}\sim {\rm L}^{z_{exp}} &;\; \tau_{int}\sim {\rm L}^{z_{int}}
  \end{array}, \end{equation}
 for large enough ${\rm L}$,
yielded the dynamical critical exponents listed in table \ref{tab-exponents}.
 Figures \ref{fig:TX0l},\ref{fig:TX0s}  are both
from the decoupled Ising model point $X_{0}$.
 In table \ref{tab-exponents} we list values of $z$
obtained from the power law fits of fig. \ref{fig:TX0l}. None the less,
 comparing these fits to the $\log {\rm L}$ fits of fig. \ref{fig:TX0s}, we
feel
that the latter ones are better and that the true value is $z=0$ for all
four autocorrelation times $\tau$.
At $X_{0}$ ( and all along the $Z=X^{2}$ or $L=0$ line )  our algorithm is
identical
with the Wolff 1C algorithm for the Ising model. Our results for $z_{int}$
from the fits in fig. \ref{fig:TX0l} are
consistent with those of Wolff \cite{Wol:comp}. A $\log {\rm L}$ behaviour has
been measured for $\tau_{exp,M}$ and $\tau_{int,M}$ for SW
dynamics in  \cite{HB} .

 At the point $X_{4}$ , the 4-state Potts
critical point (and all along the $X=Z$ line), our algorithm
is again identical with  the Wolff 1C algorithm for the 4-state Potts model.
Our result for $z_{int,E}$,
obtained from the fit in figure \ref{fig:ZcompXi} is consistent with the
result
of ref. \cite{Li:Sokal} for SW dynamics. This result can be added to the
accumulating evidence indicating: $z_{SW}=z_{Wolff}$ for $d=2$.

 The continuous variation of $z$ along the AT critical line can be
explicitly seen in figures \ref{fig:TexpE} and  \ref{fig:ZcompXi} where we
plot $\tau_{exp,E}$ and $\tau_{int,E}$ for
the 5 points $X_{i}$. In fig. \ref{fig:ZcompXi} we also plot results from
simulations at $X_{2}$
using the Metropolis method and the `naive' SW method described in sec.
\ref{sec:naive}. For the Metropolis method, we measured a value of
$z_{int,E}=1.64(8)$ , which should be compared with $z_{int,E}=.542(8)$
using our method. The `naive' SW method almost froze the
whole lattice, into a single cluster, the size of which was almost
independent of the lattice size L. For example, the average cluster
size for ${\rm L}=32$ was $\frac{\langle c\rangle}{{\rm L}^{2}}=.868(2)$ ,
as compared with
 $\frac{\langle c\rangle}{{\rm L}^{2}}=.481(1)$ using our algorithm. We
found
determination of $\tau$ for the `naive' SW method of lattice sizes ${\rm
L}>32$
so  time consuming that it was impractical. The advantage of
our method is clearly demonstrated. The three
methods yielded the same results for the static observables (within errors).

 The main question we wish to answer is whether the Li-Sokal bound is
fulfilled and whether it is sharp. In fig. \ref{fig:compZXiCv} we compare,
for the 5 points $X_{i} \;i=0\ldots4$, the exact values of
$\frac{\alpha}{\nu}$ and our
estimated values of $\frac{\alpha}{\nu}$, $z_{int,\chi}$ and $z_{exp,\chi}$.
 We see that the rise of our estimated values of $z$ follows that of
$\frac{\alpha}{\nu}$, from the decoupled Ising point $X_{0}$ to the 4-state
Potts point $X_{4}$.
 Except for the point $X_{4}$ the Li-Sokal bound is fulfilled with
respect to $(\frac{\alpha}{\nu})_{exact}$.
 The anomalously low values for $z$ at $X_{4}$ are probably caused by
 the multiplicative logarithmic correction for $C$ described in the
discussion of the specific heat results. This explanation has been suggested
by Li and Sokal \cite{Li:Sokal} to explain the low value of $z_{SW}$ for
this model.
{}From table \ref{tab-exponents} we see that $z\geq(\frac{\alpha}{\nu})_
 {estimate}$ with only one
exception at the point $X_{1}$ (see fig. \ref{fig:TX1}).
The fact that $z_{int,\chi}<
(\frac{\alpha}{\nu})_{estimate}$ could be due to a difference in
the finite size corrections or in the corrections to scaling.
$\tau_{int,\chi}$
is probably less influenced by one or both of these two factors, in
comparison with the large difference between $(\frac{\alpha}{\nu})_{
estimate}$ and $(\frac{\alpha}{\nu})_{exact}$. At the point $X_{3}$, where
the difference between $(\frac{\alpha}{\nu})_{estimate}$ and
$(\frac{\alpha}{\nu})_{exact}$ is the smallest, no anomalies in the Li-Sokal
bound occur (see fig. \ref{fig:TX3}).

We conclude that the Li-Sokal bound is fulfilled in a moderately sharp
manner. The smallest $z$ (which is $z_{int,\chi}$) fulfills
$(\frac{\alpha}{\nu})_{estimate}\leq z_{int,\chi} \leq
(\frac{\alpha}{\nu})_{estimate}+0.1$ (not including the anomalies
of the models $X_{4}$ and $X_{1}$ discussed above). Note that
we are comparing estimated values of $z$ with \underline{estimated} values of
$\frac{\alpha}{\nu}$ . This is the correct comparison to make, assuming that
$\tau$ and $C$ have similar finite size corrections and similar
corrections to scaling.
%-------------------------------------------------------------------------

%\include{sum}
\section{ Summary and Discussion }

 The correct identification
of the basic excitations of the model lead, along with the guidelines of
a general scheme for cluster algorithms, to the
construction of a cluster algorithm for the
AT model. The algorithm was shown to be identical with the one obtained by
embedding Ising spins into the AT model. Our algorithm is suitable for
spatially varying coupling constants under the restriction $L^{k,j}\leq
K^{k,j}_{\sigma},K^{k,j}_{\tau}$ everywhere on the lattice.
 We are currently carrying out intensive simulations of
a random-bond version of the AT model.

The dynamical behaviour of the cluster algorithm was examined
on the AT critical line.
 Critical slowing down of the algorithm ( $0\leq z< 1.$ ) was
found to be significantly smaller than that of the standard
Metropolis method.
A continuous variation of the dynamical exponent $z$ along the AT critical
line was seen, along with continuous variation of the static exponent
$\frac{\alpha}{\nu}$. The Li-Sokal bound $z\geq \frac{\alpha}{\nu}$, that
was proved only for $q$-state Potts models with SW dynamics, is satisfied
for   the AT model with single cluster dynamics. The bound is
moderately sharp.
  Another new result is $z_{Wolff}=z_{SW}$ for the 4-state Potts model,
which can be added to the accumulating results indicating that in two
dimensions $z_{Wolff}=z_{SW}$ .

\acknowledgments
We would like to thank Robert Swendsen and Ulli Wolff for helpful
discussions.
This research has been supported by the US-Israel Bi-national Science
Foundation (BSF).

\appendix
\section*{Ising Embedding} \label{app:embed}
Our algorithm can also be seen from a totally different point of view;
 as an example of an embedding
algorithm\cite{Wol:1C,embed:1}. The main idea is to embed into the AT
model an Ising model of space dependent couplings $J_{jk}$ and simulate it
 using the SW or Wolff procedure for the Ising model. To be explicit,
consider the Hamiltonian (\ref{eq:AT1}), and take the $\tau$ variables as
fixed, so we can write
   \begin{equation}
{\cal H}={\cal H}_{1}+{\cal H}_{2}=-\sum_{<k,j>}(K_{\sigma}+
 L\tau_{k}\tau_{j})\sigma_{k}\sigma_{j}-\sum_{<k,j>}K_{\tau}\tau_{k}\tau_{j} .
   \label{eq:AT-Is}  \end{equation}
${\cal H}_{2}$ represented by the second sum is a constant, and remembering
that we chose $L<K_{\sigma},
K_{\tau}$, \ ${\cal H}_{1}$ is a ferromagnetic Ising model in the $\sigma$
variables with couplings $J_{jk}=K_{\sigma}+L\tau_{k}\tau_{j}$. Simulating this
Ising model with any procedure that will maintain
detailed balance with respect to ${\cal H}_{1}$ and will not change the
value of ${\cal H}_{2}$, will also maintain
detailed balance with respect to ${\cal H}$. So we can use for example the
SW or Wolff procedure for the Ising model, explained in sec.
\ref{sec:general scheme}. This by itself will maintain detailed balance but
will not be ergodic since the $\tau$ variables will not be updated.
Obviously to update the $\tau$ variables the same process should be repeated
, holding the $\sigma$ variables fixed and simulating an Ising Hamiltonian
with the $\tau$ variables. To summarize, a possible procedure would go as
follows: Choose at random whether to embed into the AT Hamiltonian an
Ising Hamiltonian in the $\sigma$ spins or in the $\tau$ spins. Pick a random
site in the lattice, grow a cluster of $\sigma$ or $\tau$ spins using
the Wolff (1C) procedure using the Ising Hamiltonian and flip it.
As we will now show, this process ( we'll call it for shortness IE -- Ising
embedding) is exactly equivalent to the AT algorithm (ATA) we suggested in
 section \ref{sec:ATC} .

A first  argument  which is really sufficient goes as follows. Suppose we
perform the IE freeze-delete scheme for the $\sigma$ spins for the whole
lattice, viewing the $\tau$ spins as constant. We could say we've generated
a new Hamiltonian of noninteracting clusters of $\sigma$ spins and which
assigns infinite energy to any configuration which differs from the current
configuration of $\tau$ spins. Now suppose we perform the ATA freeze-delete
scheme for the whole lattice. Then we get a new Hamiltonian of
noninteracting clusters of $\sigma$ spins and of $\tau$ spins, but if at
this stage we decide to flip only $\sigma$ spins, then in practice we've
assigned infinite energy to any configuration which differs from the current
configuration of $\tau$ spins. In practice we flip only one cluster of the
$\sigma$ spins, so what we do is identical to the Wolff (1C) version of the
IE. Since both processes maintain detailed balance using the same new
Hamiltonian they must do so using the same probabilities (the general
scheme in sec. \ref{sec:general scheme} shows a one to one correspondence
between probabilities and the new Hamiltonian.),
which completes our argument that the two methods are actually identical.

One can, of course, check that the probabilities of the two procedures are
the same. For example, denote the probability to delete a bond between
$\sigma$ spins in ATA as $P_{d,\sigma}^{ATA}(u)$.  for $u_{1}$
for example, From fig. \ref{fig:scheme}:
   \begin{equation}
P_{d,\sigma}^{ATA}(u_{1})=P_{d}(u_{1})+q_{2}=e^{E_{1}-E_{2}}=
e^{-2(K_{\sigma}+L)}. \label{eq:del ATA}      \end{equation}
Now with the Ising Embedding algorithm, according to (\ref{eq:swd1}):
   \begin{equation}
P_{d,\sigma}^{IE}(u) =e^{-(K_{\sigma}+L\tau_{k}\tau_{j})(\sigma_{k}\sigma_{j}
+1)} ,    \label{eq:del IE}\end{equation}
so $P_{d,\sigma}^{IE}(u_{1})=P_{d,\sigma}^{ATA}(u_{1})$. This check can be
carried out for all $u_{i}$.

% now the references. delete or change fake bibitem. delete next three
%   lines and directly read in your .bbl file if you use bibtex.
%\begin{references}
%\bibitem{tag} Fake bibitem.
%\end{references}

%  \include{figs1}
\begin{figure}
\caption{Phase diagram of the Ashkin Teller (Z(4) ) model.}
\label{fig:AT phase}
\end{figure}

\begin{figure}
\caption{States of spins at a pair of
nearest-neighbor sites $j,k$.Each state $u_{i}$ represents one out of four
states with the same energy $E_{i}$.}
 \label{fig:states} \end{figure}

\begin{figure}
\caption{ Freeze/delete scheme for the AT model. The four states of two
neighboring sites are denoted in the left column; Four freeze/delete
operations are denoted in the top row, and the respective probabilities are
denoted in the table. A frozen bond is denoted by a double thick line,
while a deleted bond is denoted by a blank.}
\label{fig:scheme} \end{figure}

\begin{figure}
\caption{ `Naive' SW freeze/delete scheme for the AT model. The four states
of two neighboring sites are denoted in the left column; Four freeze/delete
operations are denoted in the top row, and the respective probabilities are
denoted in the table. A frozen bond is denoted by two thick lines, while a
deleted bond is denoted by a blank. }
\label{fig:SW AT} \end{figure}

\begin{figure}
\caption{ A log-log plot of specific heat vs. linear lattice size at the
five critical models $X_{0}\ldots X_{4}$. The solid lines are the linear
fits. the critical exponents $\frac{\alpha}{\nu}$ are given in table 1.}
\label{fig:Cv graph} \end{figure}

\begin{figure}
\caption{ The $\ln$ of the time auto-correlation function of the
susceptibility $\phi_{\chi}(t)$, measured for the $X_{2}$ model at L=128.
The unit of time is a single
MC step or a single cluster flip. The vertical lines are the error bars. The
exponential fit is also shown.}
\label{fig:fi(t)} \end{figure}

\begin{figure}
\caption{ Log-log plots of auto-correlation times $\tau_{\chi}$ and the
specific
heat $C$ as a function of linear lattice size ${\rm L}$, at the decoupled
Ising point $X_{0}$. The values of $z$ listed in table 1 are the slopes
of the fits.}
\label{fig:TX0l} \end{figure}

\begin{figure}
\caption{ Semi-log plots of auto-correlation times $\tau_{\chi}$ and the
specific
heat $C$ as a function of linear lattice size ${\rm L}$, at the decoupled
Ising point $X_{0}$. This fit seems to be better than the log-log fit,
yielding values of $z=\frac{\alpha}{\nu}=0$.}
\label{fig:TX0s} \end{figure}

\begin{figure}
\caption{ Log-log plots of auto-correlation times $\tau_{\chi}$ and the
specific heat $C$ as a function of linear lattice size ${\rm L}$, at the
 point $X_{1}$. The values of $z$ listed in table 1 are the slopes
of the fits.}
\label{fig:TX1} \end{figure}

\begin{figure}
\caption{ Log-log plots of auto-correlation times $\tau_{\chi}$ and the
specific heat $C$ as a function of linear lattice size ${\rm L}$, at the
 point $X_{3}$. The values of $z$ listed in table 1 are the slopes
of the fits.}
\label{fig:TX3} \end{figure}

\begin{figure}
\caption{ A log-log plot of $\tau_{exp,E}$ vs. linear lattice size at the
five critical models $X_{0}\ldots X_{4}$. The solid lines are the linear
fits. the critical exponents $z_{exp,E}$ are given in table 1.}
\label{fig:TexpE} \end{figure}

\begin{figure}
\caption{ Log-log plots of $\tau_{int,E}$
 as a function of linear lattice size ${\rm L}$, at all 5 points
$X_{i}\;i=0\ldots
4$. The continuous variation of the slope $z$ along the AT critical line
is easily seen.
Metropolis results and `naive' SW results for $X_{2}$ are also plotted.}
\label{fig:ZcompXi} \end{figure}

\begin{figure}
\caption{ Comparison of $z$ and $\frac{\alpha}{\nu}$ at the 5 points
$X_{i}\;\; i=0\ldots4$.
The line denotes the exact value of $\frac{\alpha}{\nu}$, while our
estimated values for the five models are denoted by full circles. values of
$z_{int,\chi}/z_{exp,\chi}$  are denoted by empty circles/ crosses.}
\label{fig:compZXiCv} \end{figure}

%********************************************************************

 \begin {table}
\caption{  Results from the AT critical line. The errors in parentheses are
only the statistical errors of the fit in our fitting interval. They do not
include systematic errors stemming from finite size effects and corrections
to scaling.}
\label{tab-exponents}
\begin{tabular}{llllllll}
      & $\frac{\gamma}{\nu}$&$\frac{\alpha}{\nu}$& $z_{int,E}$ & $z_{exp,E}$
& $z_{int,\chi}$ & $z_{exp,\chi}$&$\frac{\alpha}{\nu}_{exact}$\\ \hline
$X_{0}$& 1.751(4)& .23(1)\,/\,0& .26(3)& .23(2)&.13(1)& .267(4)&  0   \\
$X_{1}$& 1.751(1)& .38(2) & .396(5)& .40(3)& .273(3)& .47(2) & .2096\\
$X_{2}$& 1.752(3)& .542(8) & .61(1)& .68(3)& .532(5)& .62(1) & .4182\\
$X_{3}$& 1.763(6)& .655(13)& .719(2)& .74(2)& .717(8)& .72(3) & .6383\\
$X_{4}$& 1.752(4)& .747(3) & .92(1)& .99(5)& .99(3)& .94(3) &  1   \\
 \end{tabular}
\end{table}


\begin{thebibliography}{99}
\bibitem{q}  Actually the algorithm is for the q-state Potts model.
\bibitem{L<K} If $L=K_{\sigma}$ or $L=K_{\tau}$ then $E_{4}=E_{3}$
or $E_{4}=E_{2}$. We do not discuss here the behaviour of our algorithm in
such a case.
\bibitem{coup}
 From now on we'll absorb $\beta$ into
the coupling constants of the Hamiltonian. We'll also omit the $l$
dependence of all constants of the algorithm since we are considering, for
the time being, constant $K_{\sigma},K_{\tau}$ and $L$. Generalization to
spatially
varying couplings is trivial as long as $K_{\sigma},K_{\tau}\geq L$
everywhere.
\bibitem{energy} A factor of $k_{B}T$ has been omitted,
but is irrelevant since all measurements are at a single temperature.
\end{thebibliography}
\end{document}